**Comments on "*Fundamental nature of the self-field critical current in superconductors*", arXiv:2409.16758.**

A. Goyal[1*], R. Kumar[1], A. Galluzzi[2] and M. Polichetti[2*]
[1]Laboratory for Heteroepitaxial Growth of Functional Materials & Devices, Department of Chemical & Biological Engineering, State University of New York (SUNY) at Buffalo, Buffalo, NY, USA.
[2]Laboratorio "LAMBDA" – Dipartimento di Fisica, Università di Salerno and CNR-SPIN Unità di Salerno, Fisciano (SA), Italy.
*Corresponding authors: A. Goyal (email: agoyal@buffalo.edu) and M. Polichetti (email: mpolichetti@unisa.it).

*Abstract*: We provide comments on article "*Fundamental nature of the self-field critical current in superconductors*" by Talantsev and Tallon, arxiv:2409.16758 and https://papers.ssrn.com/sol3/papers.cfm?abstract_id=4978589 related to the predictions of the key equation proposed in article "*Universal self-field critical current for thin-film superconductors*," by Talantsev and Tallon, Nat. Commun. 6 (2015) 7820. We respond to claims and assertions made in these articles and also show that the so-called "*Universal self-field critical current*" and the "*fundamental limit*" for self-field $J_c$ suggested by Talantsev and Tallon, has been proven incorrect experimentally and this invalidates the key claims and findings stated in both these articles.

Potential large-scale applications of superconductors, in particular, energy generation via commercial nuclear fusion has generated enormous interest world-wide due to potentially transformative impacts on addressing the world's energy needs [1]. Recently, Talantsev and Tallon reached out to us and requested raw data related to magnetization for the $J_{c,mag}$ reported by Goyal et al. [1]. This data was promptly, and without any delay provided to them in the form of ΔM, or volume magnetization from three of figures reported in Goyal et al. along with sample dimensions of width, length and thickness [1]. Upon receiving this data, they calculated $J_{c,mag}$. They communicated to us that one of them calculated $J_{c,mag}$ to be a factor of 10 lower than that reported by Goyal et al. [1] and the other calculated $J_{c,mag}$ to be higher than a factor of $10^3$ than that reported by Goyal et al. [1] and also provided us details of these calculations arguing that both were perfectly valid and correct methods to calculate $J_{c,mag}$ from ΔM values. They stated the above ambiguity in calculation of $J_{c,mag}$, to request moment data from us, not appreciating that ΔM, or volume magnetization is merely the raw moment divided by the sample volume.

In response to their email, we detailed to them our step-by-step methodology used to calculate $J_{c,mag}$ in Goyal et al. [1], and which is also briefly summarized in the article titled "*Expression with self-consistent magnetic units for calculation of critical current density using the Bean's Model*" [2]. We also immediately began exploring in great detail how the calculation of $J_{c,mag}$ using the Bean's model had been done in the literature and to make sure nothing was done incorrectly. In doing so, when we found the first reference where the magnetization and sample dimensions were provided, allowing us to reproduce the calculations to obtain the $J_{c,mag}$ calculated using equation 2 reported in Reference 2, and also shown below for completeness,

$$J_{c,mag} = \frac{2 \times \Delta M}{w \times \left(1 - \frac{w}{3b}\right)} \qquad (1)$$

we sent a request to Nat. Commun. for retraction of our paper. While this request was sent in mid-September, due to editorial/journal processing time the actual formal retraction by Nat. Commun. occurred on October 23$^{rd}$, 2024 [3]. Prior to this retraction, a request for correction of some typographical errors was sent to Nat. Commun. in August, 2024, very shortly after publication of the article in August. However, since it takes the journal many weeks to do this, and since our request for retraction of our paper was made prior to these corrections being officially made, again for editorial reasons Nat. Commun. halted the corrections and stated that these requested corrections from August will appear with the official retraction note, as they did [3].

Since in the final analysis, transport $J_c$ is of interest for applications, we plan to conduct a full measurement of $J_c$(H,T) transport from 4.2K to 77K in fields upto 7T (similar to field and temperature range in Goyal et al. [1]) and then compare to $J_{c,mag}$ using the equation above, and will report this at a future date.

Subsequent to us providing Tallon the step-by-step methodology used to calculate $J_{c,mag}$ in Goyal et al. [1], a week or so later, we received a first version of the article published afterwards in ArXiv in September, 2024, from Tallon[*]. This article revealed that besides merely alerting us to a possible issue in calculation of $J_{c,mag}$ in Goyal et al [1], a key interest was in pointing out the importance of their prior work, also published in Nat. Commun. and shown in reference 4. In the following sections, we comment on some aspects of this article published in arXiv and SSRN [5] and stemming directly from the theoretical predictions of the key equations in the Nat. Commun. paper "*Universal self-field critical current for thin-film superconductors*," by Talantsev and Tallon [4].

The original arXiv article sent to us by Talantsev and Tallon (prior to submission to arXiv and SSRN) claimed that per their theoretical analysis reported in reference 4, the "*fundamental limit*" of attainable self-field $J_c$ was $J_c$(sf,$T \rightarrow 0$ K) ≤ 45 MA/cm$^2$. This was calculated using the reported value of the London penetration depth ($\lambda(T \rightarrow 0\ K) \approx 105\ nm$, in the *b* direction) by Kiefl et al. [6]. In that article, they also stated that this self-field $J_c$ has now been achieved by all major 2G-wire manufacturers worldwide. In response to their article, we pointed out to them three highly regarded and published papers reporting self-field $J_c$ values which substantially exceeded this so-called "*fundamental limit*" of $J_c$(sf,$T \rightarrow 0$ K) ≤ 45 MA/cm$^2$ [7,8,9].

The first paper by Xu et al. [7], reported a transport $J_c$ (sf, 4.2K) ~ 55 MA/cm$^2$ at 4.2K for H//c for a thin-film superconductor of composition 15 mol.% Zr-added (Gd, Y)-Ba-Cu-O. This $J_c$ (sf, 4.2K) was ~ 22% higher than the so-called "*fundamental limit*" of 45 MA/cm$^2$ claimed by them initially. If the measurement was made at lower temperatures, approaching 0K, the $J_c$ would have increased manifold.

---

[*] This version of the article (with the so-called "*fundamental limit*" of attainable self-field $J_c$ stated to be $J_c$(sf,$T \rightarrow 0$ K) ≤ 45 MA/cm$^2$) was also widely disseminated to the superconductor scientific community in September 2024, including to the Materials Sub Committee of the 17$^{th}$ European Conference on Applied Superconductivity (EUCAS) to be held 21$^{st}$-25$^{th}$ September 2025, Porto, Portugal.

The second paper, by Stangl et al. [8], reported a $J_c$ (sf, 5K) of ~ 90 MA cm$^{-2}$ in an undoped, YBa$_2$Cu$_3$O$_{7-\delta}$ (YBCO) thin film. This $J_c$ (sf, 5K) is 100% higher than the so-called "*fundamental limit*" of 45 MA/cm$^2$ claimed by them initially. Again, if the measurement was made at lower temperatures, approaching 0K, the $J_c$ would have increased manifold.

The third paper, by Miura et al. [9], reported both a transport and magnetization $J_c$ (sf, 4.2K) of 130 MA cm$^{-2}$ in a (Y,Gd)Ba$_2$Cu$_3$O$_{7-\delta}$ (YBCO) thin film with BaHfO$_3$ additions. This $J_c$ (sf, 5K) is 188% higher than the so-called "*fundamental limit*" of 45 MA/cm$^2$ claimed by them initially. Again, if the measurement was made at lower temperatures, approaching 0K, the $J_c$ would have increased manifold.

It was pointed out to Tallon and Talantsev that these experimental results invalidate their theory and show that the so-called "*fundamental limit*" to $J_c$(sf) reported in their paper (Ref. 4), *Nat. Commun.* **6**, 7820 (2015), is incorrect and the real $J_c$(sf) "*fundamental limit*" is significantly higher. Unfortunately, they responded by questioning the validitity of the results reported in papers cited above which exceeded their so-called "*fundamental limit*" to $J_c$(sf).

Subsequently, in an attempt to address our comments mentioned above, based on a value for an effective penetration length $\lambda_{\text{eff}} = \sqrt{\lambda_a \lambda_b} = 91\ nm$ reported in the paper by Pereg-Barnea et al. (Reference 4 in arXiv and SSRN articles [5], or reference 7 in the paper by Kiefl et al. [6]), Talantsev and Tallon modified the *"fundamental limit" of self-field $J_c$ from ~ 45 MA/cm$^2$ to be ~ 78 MA/cm$^2$* as reported in the article submitted to arXiv and SSRN [5], and claimed that this "*has been reported by Stangl et al., for overdoped pulsed-laser-deposited films*". Nevertheless, the value of $\lambda_{\text{eff}}$ used by Talantsev and Tallon to increase the value of the *"fundamental limit"* has never been confirmed, as correctly indicated by them in the arXiv and SSRN articles [5], despite being reported in a paper published in 2004. Moreover, the value relates to a Gd$_x$Y$_{1-x}$Ba$_2$Cu$_3$O$_{6+y}$ single crystal, whereas in the paper by Stangl et al. [8] the reported value of 90 MA/cm$^2$ refers to a YBCO film without Gd, at a temperature higher than T=0, with larger values of both the characteristic lengths (see Table 1 in ref [8]), and in particular of $\lambda_{ab}$, which will correspond to a "*fundamental limit*" of ~ 44 *MA/cm$^2$*, substantially lower than 78 MA/cm$^2$, calculated using their primary equation (*Eq. 4*) in the Nat. Commun. paper published in 2015 [4] and the primary equation (*Eq. 1*) in arXiv and SSRN articles [5]. We show below this calculation of the so-called "*fundamental limit*" to self-field $J_c$ based on the primary equation (*Eq. 4*) of Talantsev and Tallon reported in Nat. Commun. [4], using the thermodynamic parameters experimentally measured and reported by Stangl et al. of the London penetration depth, $\lambda_{ab}$, and coherence length, $\xi$ of:

$\lambda = 112\ nm = 112 \cdot 10^{-9}\ m$, and $\xi = 1.6\ nm = 1.6 \cdot 10^{-9}\ m$ (for the sample with $J_c(5K) = 90\ \frac{MA}{cm^2}$),

$$Jc = 1.309 \cdot 10^{-10} Am \times \frac{\ln\left(\frac{112 \cdot 10^{-9}\ m}{1.6 \cdot 10^{-9}\ m}\right) + 0.5}{(112 \cdot 10^{-9})^3\ m^3} = 1.309 \cdot 10^{-10} Am \times \frac{4.25 + 0.5}{1.405 \cdot 10^{-21}\ m^3} =$$
$$= 1.309 \cdot 10^{-10} \times 3.38 \cdot 10^{21} \left(\frac{Am}{m^3}\right) \cong 4.4 \cdot 10^{11}\ \frac{A}{m^2} = 4.4 \cdot 10^7\ \frac{A}{cm^2} = 44\ \frac{MA}{cm^2}$$

The experimentally reported $J_c$ by Stangl et al. is 100% larger (90 $\frac{MA}{cm^2}$) than this so-called "*fundamental limit*" to self-field $J_c$.

Talantsev and Tallon claim that their work in References 4 and 5 establishes an upper limit for the highest achievable, self-field, <u>transport</u> critical current in thin-film superconductors (see page 1, second paragraph, reference 5), and that this *"fundamental limit" is ~ 78 MA/cm$^2$*. However, there is NO clear data supporting this result and, indeed, there are at least two clear experimental self-field $J_c$ reports in literature [8, 9] if not three experimental self-field $J_c$ reports [7,8,9] indicating that the theoretical analysis reported by Talantsev and Tallon [4] is *incorrect* and *invalid*.

In reference 5, Talantsev and Tallon cite the formula of Gyorgy et al. [10]. As explained in detail in reference 2, Talantsev and Tallon's analysis is incorrect. From Gyorgy *et al.*, it is clear that if ΔM is expressed in ***gauss*** (not in *emu/cm$^3$* as stated by Talantsev and Tallon), and sample dimensions in cm, the resulting current density is in A/cm$^2$, *without* any other conversion of units. This can be checked also by considering the data and specific examples reported in Gyorgy *et al*. In order to use Gyorgy *et al*. with the correct magnetic units, if ΔM is in emu/cm$^3$, it has to be converted to gauss by multiplying it by 4π, per standard tables of magnetic unit conversions as shown in databases of NIST and IEEE [11, 12], yielding a $J_c$ higher by a factor of 12.56 than that calculated using the equation (1) reported above [2]. Moreover, in reference 5, Talantsev and Tallon affirm that in the equation reported in Gyorgy et al. the multiplicative term 20 is not dimensionless but its units are [A cm$^2$ emu$^{-1}$]. This is incorrect, as explained in detail in references 2 and 13 of this article.

In Fig. 2b of Talantsev and Tallon's arXiv paper (reference 5 of this article), Talantsev and Tallon compare the $J_c$ of two samples from S-Innovations - $J_{c,mag}$ for a 2 μm thick film compared to a $J_{c,transport}$ for a 2.8 μm thick film. This makes no scientific sense whatsoever. We expect $J_{c, transport}$ to be typically higher than $J_{c,mag}$ for the same film. Only data for a film of the *SAME composition*, fabricated at the *SAME time* and of the *SAME thickness* for both transport and magnetic $J_c$ measurements makes any scientific sense to compare. Else, it is easy to find films of different thicknesses fabricated at different times under different conditions to match a $J_{c, mag}$ with $J_{c, transport}$. Even in the plot shown in Fig. 2b, $J_{c, transport}$ for a much thicker film is 10-20% higher than $J_{c,mag}$ depending on the measurement temperature. Else, Figure 2b has no relevance. Films of different thickness are expected to have different $J_c$'s. Films of the same nominal composition and same thickness can also vary in $J_c$ from one deposition run to another. Hence, it is important that both transport and magnetic $J_c$ measurements need to be performed on films of the *SAME composition*, having the *SAME thickness* and made at the *SAME time*. Fig. 2c is of value only if $J_{c,mag}$ is compared to $J_{c,mag}$ using various methods including the Talantsev and Tallon's methodology for calculating $J_{c,mag}$ higher by a factor 10$^3$ than that reported by Goyal et al. [1].

Talantsev and Tallon point out that they have identified other "*10-fold mistakes*" in Goyal et al. [1]. In this article, they point to error in calculations of microstrain using the Williamson-Hall method even though it had already been pointed out to them that the calculations reported in the paper pertaining to microstrain estimation using the Williamson-Hall method are absolutely accurate as reported in figures corresponding to this data, i.e. Figs. 3c and 4c of Goyal et al. In addition, it was explicitly pointed out to Talantsev and Tallon that the typographical error in the text of the paper that they point to, was included as part of the typographical corrections requested to be made by Nat. Commun. in August 2024 and that these had not yet been made by the journal. As pointed out previously, the journal decided to halt these corrections once we made an article retraction request and stated that these will instead be included together with the retraction note (which was published on Oct. 23rd, 2024 [3]).

Given that Talantsev and Tallon were explicitly informed as mentioned above, their insistence on including the statement about "*other 10-fold mistakes*" in reference 5, reveals the need to have additional support, albeit from mere typographical errors, to make the case for the key message in their article stronger.

The final and most important is the incorrect conclusion drawn by Talantsev and Tallon in the last paragraph - "*we have also confirmed the validity of our primary equation (Eq. 1) for the self-field critical current density in superconductors*". This is an incorrect statement since at least two outstanding and highly cited publications in the field (Stangl et al. [8] and Miura et al. [9]) have reported a significantly higher self-field $J_c$ than this so-called "*fundamental limit*" for self-field $J_c$. Such an incorrect statement <u>harms the field</u> as some wire manufacturers may rest on this statement that they have almost approached the "*fundamental limit*" to $J_c$ and not strive to achieve the performance already shown possible by the excellent work of Stangl et al. [8] and Miura et al. [9]. The theoretical analysis by Talantsev and Tallon reported in References 4 and 5 has been proven incorrect experimentally and this also invalidates their original Nat. Commun. paper published in 2015 [4].

**Acknowledgements:**

R. Kumar and A. Goyal were supported by ONR Grant No. N00014-21-1-2534.

**References:**

1. Goyal, A., Kumar, R., Yuan, H. *et al.* Significantly enhanced critical current density and pinning force in nanostructured, (RE)BCO-based, coated conductor. *Nat Commun* 15, 6523 (2024). https://doi.org/10.1038/s41467-024-50838-4.
2. Polichetti, M., Galluzzi, A., Kumar, R. & Goyal, A. Expression with self-consistent magnetic units for calculation of critical current density using the Bean's Model, http://arxiv.org/abs/2410.09197.
3. Goyal, A., Kumar, R., Yuan, H. *et al.* Retraction Note: Significantly enhanced critical current density and pinning force in nanostructured, (RE)BCO-based, coated conductor. *Nat Commun* 15, 9140 (2024). https://dx.doi.org/10.1038/s41467-024-53580-z.
4. E.F. Talantsev, J.L. Tallon, Universal self-field critical current for thin-film superconductors. Nat Commun. 6 (2015) 7820.
5. Talantsev and Tallon, Fundamental nature of the self-field critical current in superconductors, https://arxiv.org/abs/2409.16758 and https://papers.ssrn.com/sol3/papers.cfm?abstract_id=4978589.
6. Kiefl, R. F. et al. Direct measurement of the London penetration depth in $YBa_2Cu_3O_{6.92}$ using low-energy μSR. Phys Rev B 81, 180502 (2010).
7. A. Xu, L. Delgado, N. Khatri, Y. Liu, V. Selvamanickam, D. Abraimov, J. Jaroszynski, F. Kametani, D. C. Larbalestier; Strongly enhanced vortex pinning from 4 to 77 K in magnetic fields up to 31 T in 15 mol.% Zr-added (Gd, Y)-Ba-Cu-O superconducting tapes. *APL Mater.* 1 April 2014; 2 (4): 046111. https://doi.org/10.1063/1.4872060.
8. Stangl, A., Palau, A., Deutscher, G. *et al.* Ultra-high critical current densities of superconducting $YBa_2Cu_3O_{7-\delta}$ thin films in the overdoped state. *Sci Rep* 11, 8176 (2021). https://doi.org/10.1038/s41598-021-87639-4.
9. Miura, M., Tsuchiya, G., Harada, T. *et al.* Thermodynamic approach for enhancing superconducting critical current performance. *NPG Asia Mater* 14, 85 (2022). https://doi.org/10.1038/s41427-022-00432-1.
10. E. M. Gyorgy; R. B. van Dover; K. A. Jackson; L. F. Schneemeyer; J. V. Waszczak, Anisotropic critical currents in $Ba_2YCu_3O_7$ analyzed using an extended Bean model, *Appl. Phys. Lett.* 55, 283–285 (1989).
11. https://www.nist.gov/system/files/documents/pml/electromagnetics/magnetics/magnetic_units.pdf.
12. https://ieeemagnetics.org/files/ieeemagnetics/2022-04/magnetic_units.pdf.
13. Polichetti, M., Galluzzi, A., Kumar, R. & Goyal, A. Comments on "Procedures for proper validation of record critical current density claims" by Chiara Tarantini and David C. Larbalestier, https://arxiv.org/abs/2411.16685.